\documentstyle[twocolumn,aps,psfig]{revtex}
\begin{document}
\draft
\title{Viscous cosmologies in scalar-tensor theories for Kasner type metrics.}
\author{Mauricio Cataldo$^{\,\,a}$ {\thanks{E-mail address: 
mcataldo@alihuen.ciencias.ubiobio.cl}},
Sergio del Campo $^{b}$ \thanks{%
E-mail address: sdelcamp@ucv.cl} and Patricio Salgado$^{c}$ {\thanks{%
E-mail address: pasalgad@udec.cl}}}
\address{$^a$
Departamento de F\'\i sica, Facultad de Ciencias, Universidad del
B\'\i o-B\'\i o, Avda. Collao 1202, Casilla 5-C, Concepci\'on,
Chile. \\ $^b$ Instituto de F\'\i sica, Facultad de Ciencias
B\'asicas y Matem\'aticas, Universidad Cat\'olica de Valpara\'\i
so, Avenida Brasil 2950, Valpara\'\i so, Chile. \\ $^c$
Departamento de F\'\i sica, Facultad de Ciencias F\'\i sicas y
Matem\'aticas, Universidad de Concepci\'on, Casilla 4009,
Concepci\'on, Chile.} \maketitle
\begin{abstract}
{\bf {Abstract:}} In a viscous Bianchi type I metric of the Kasner
form, it is well known that it is not possible to describe an
anisotropic physical model of the universe, which satisfies the
second law of thermodynamics and the dominant energy condition
(DEC) in Einstein's theory of gravity.

We examine this problem in scalar-tensor theories of gravity. In
this theory we show that it is possible to describe the growth of
entropy, keeping the thermodynamics and the dominant energy
condition. \vspace{0.5cm}

PACS number(s): {98.80.Hw, 98.80.Bp}
\end{abstract}
\smallskip\
\section{Introduction}
The standard Friedmann-Robertson-Walker (FRW) model has attracted
considerable attention in the relativistic cosmology literature,
since the FRW model is in accordance with the large scale spatial
homogeneity and isotropy of the observable universe. Perhaps, one
of the most important characteristics of this model is, as
predicted by inflation~\cite{Guth}, the flatness, which agrees
with the observed  cosmic microwave background radiation.

One can consider as source for FRW models not only a perfect
fluid, without dissipative viscous processes, but also a bulk
viscosity, which is compatible with the isotropy of the
universe~\cite{Weinberg,Johri}. Therefore, the imperfect fluid
also can be considered as a source for FRW models. It is shown in
ref.~\cite{Tupper1}, for a ``tilting" four velocity with a
spacelike component, that FRW cosmologies, in particular flat
models, do not necessarily represent perfect fluid solutions. They
may correspond to exact solutions of the field equations of
viscous fluid in which bulk and shear viscosity are considered. In
this case, the heat conduction vector $q_{\mu}$ is not zero.

In the early universe the kinds of matter fields are uncertain.
The presence of anisotropy at early times is a very natural idea
to explore, as an attempt to explain, among other things, the
local anisotropies that we observe today in galaxies, clusters and
superclusters. Thus, at early time, it seems appropriate to assume
a geometry that is more general than merely the isotropic and
homogeneous FRW geometry. Even though the universe, on a large
scale, seems homogeneous and isotropic at the present time, there
are no observational data that guarantee the isotropy in an era
prior to the recombination. In fact, it is possible to begin with
an anisotropic universe which isotropizes during its evolution by
the damping of this anisotropy via a mechanism of viscous
dissipation. The anisotropies described above have many possible
sources: they could be associated with cosmological magnetic or
electric fields, long-wavelength gravitational waves, Yang-Mills
fields, axion fields in low-energy string theory or topological
defects such as cosmic strings or domain walls, among others (see
ref.~\cite{Barrow} and references therein).

We should mention that, in the Einstein theory of gravity, when an
anisotropic Bianchi type-I model of the Kasner form is considered,
it is not possible to describe the growth of entropy, due to the
incompatibility between the second law of thermodynamics and the
dominant energy condition~\cite{Cataldo1,Brevik1}. In this paper
we would like to explore this problem in the situation in which a
more general theory of gravity is considered: specifically,
scalar-tensor theories. In this respect we shall consider a sort
of Jordan-Brans-Dicke scalar field which couples to gravity in
presence of a viscous fluid in an anisotropic Bianchi type-I model
of the Kasner form
\begin{eqnarray}
\label{Metric} \displaystyle ds^2= dt^2 - t^{2 p_1} dx^2- t^{2
p_2} dy^2- t^{2 p_3} dz^2,
\end{eqnarray}
where $p_1$, $p_2$ and $p_3$ are three parameters that we shall
require to be constants. The expansion factors $t^{p_1}$, $
t^{p_2}$ and $t^{p_3}$ will be determined via Einstein's field
equations. The space is anisotropic if at least two of the three
$p_i$ ($i=1,2,3$) are different.

In the next section we give the definitions of the corresponding
expressions that we shall use throughout this paper. Also, in the
same section, we show that, in standard Einstein relativity,  a
viscous cosmological fluid does not permit the Kasner metric to be
anisotropic\cite{Cataldo1,Brevik1,Brevik}.  We work out the
general scalar - tensor theory in Sect.III, and go on to analyze
specific models in Sect.IV. Whereas in most cases conflicts
between anisotropy/viscosity and the dominant energy condition
still turn out to be the case, there are a few exceptions, as
exemplified in Sect.IV.C,  where the set of  given anisotropic
Kasner coefficients leads to a thermodynamically consistent
description of the viscous universe. In section V we calculate the
generation of entropy for a thermodynamically consistent
cosmological model. Finally, in section VI we conclude our work.

\section{Viscous Kasner type universe in Einstein 
theory}
The Kasner universe, in Einstein's theory (with cosmological
constant $\Lambda=0$), refers to a vacuum cosmological model given
by~(\ref{Metric}), where the numbers $p_1$, $p_2$ and $p_3$
satisfy the constraints
\begin{eqnarray}  \label{condiciones del Kasner}
p_1+p_2+p_3=1, \,\,\,\,\, p^2_1+p^2_2+p^2_3=1.
\end{eqnarray}
An anisotropic Kasner type universe can be considered to be filled
with an ideal (nonviscous) fluid which has an equation of state
$p=\rho$ (stiff matter - the velocity of sound coincides with the
speed of light), where $\rho$ is the energy density and $p$ is the
isotropic pressure. In this case only the constraint
$p_1+p_2+p_3=1$ is satisfied.

When one considers a fluid with bulk and shear viscosity
coefficients, i.e. a viscous fluid, in a Kasner type universe one
can obtain mathematically self-consistent solutions. Here, the
mass density and the isotropic pressure are proportional to
$1/t^2$, the bulk and shear viscosity coefficients are
proportional to $1/t$, and the constraints~(\ref{condiciones del
Kasner}) are not satisfied any more~\cite{Brevik}. In general the
energy-momentum tensor for a viscous fluid is given by
\begin{eqnarray}  \label{tensor con shear}
T_{\alpha \beta}= [\rho + (p-\xi \theta)] u_{\alpha} u_{\beta} -
(p-\xi \theta) g_{\alpha \beta} + 2 \eta \sigma_{\alpha \beta},
\end{eqnarray}
where $u_\alpha$, $\rho$, $p$, $\xi$ and $\eta$ are the fluid's
four velocity, the energy density, the isotropic pressure, the
bulk and shear (or coefficient of dynamic viscosity) viscosities,
respectively. The scalar expansion and the traceless shear tensor
are defined by $\theta= u^{\alpha}_{\,\, ; \alpha}$ and
\begin{eqnarray}
\sigma_{\alpha \beta} = h_{\alpha}^{\gamma} u_{(\gamma ; \delta)}
h^{\delta}_{\beta}- \frac{1}{3} \theta h_{\alpha \beta},
\end{eqnarray}
respectively. Here $h_{\alpha \beta}=g_{\alpha \beta}-u_{\alpha}
u_{\beta}$ is the projection tensor.

 From the metric~(\ref{Metric}) and the Einstein equations we get
for the shear viscosity the following expression
\begin{eqnarray}  \label{expresion 2}
\eta = \frac{1}{16 \pi G \, t} (1-p_1-p_2-p_3).
\end{eqnarray}

It could be shown from thermodynamic principles that we should
impose the conditions $\xi \geq 0$ and $\eta \geq 0$ in order to
have a positive entropy generation~\cite{MiThWe}. When these
conditions are applied in an anisotropic Kasner type model, the
parameters entering into the metric have to satisfy the bound
$p_1+p_2+p_3 \leq 1$ in order to deal with an appropriate physical
model. But, as was shown in refs.~\cite{Cataldo1,Brevik1}, it is
not possible to describe the growth of entropy in an anisotropic
model of the Kasner form while keeping the 2nd law of
thermodynamics together with the DEC.

On the other hand, we require the model to satisfy the DEC. This
becomes specified by the range $- \rho \leq P_j \leq
\rho$~\cite{Hawking}, where $ \rho$ is the energy density and
$P_j$ (with $j = x,y,z$) are the effective momenta related to the
corresponding coordinate axes. Note that both $\rho$ and $P_j$
scale as $t^{-2}$. Thus, the DEC will give some specific relations
among the Kasner parameters $p_i$ ($i=1,2,3$). In the following,
we introduce the symbols $S \equiv p_1+p_2+p_3$ and $Q \equiv
p^2_1+p^2_2+p^2_3$, following ref.~\cite{Brevik}.

The conditions $P_j \leq \rho$, with $j = x,y,z$, yield three
inequalities, which, by adding them, reduce to just one inequality
given by
\begin{eqnarray}  \label{expresion 5}
2 S (S-1) \geq 0.
\end{eqnarray}
In a similar way, from $P_j \geq - \rho$ we get the inequality
\begin{eqnarray}  \label{expresion7}
S \geq \frac{A S^2}{2},
\end{eqnarray}
where $A=3Q/S^2-1$~\cite{Cataldo1}.

 From expression~(\ref{expresion7}), we see that $S \geq 0$, since
$A \geq~0$~ \cite{Brevik,Cataldo1}. With this condition on $S$, we
obtain from expression~(\ref{expresion 5}) that necessarily $S$
should be greater than one. This conclusion yields to a negative
shear viscosity, as can be seen from expression~(\ref{expresion
2}). Therefore, from this result, we observe that the entropy in
this sort of universe decreases instead of increasing.

We should note here that the DEC, specified by the range for $P_i$
($i=x,y,z$), $- \rho \leq P_i \leq \rho$, is equivalent to the
range $-\rho \leq p \leq \rho$, where $p$ is the isotropic
pressure. Effectively, for a fluid with energy density $\rho(t)$
and principal pressures $P_{x}(t)$, $P_{y}(t)$ and $P_{z}(t)$, the
energy-momentum tensor becomes $T^{\alpha}_{\beta}=
diag(\rho,-P_{x},-P_{y},-P_{z})$~\cite{Barrow,Hawking}. Thus, from
eqs.~(\ref{Metric}) and~(\ref{tensor con shear}), and using
$u_\alpha=\delta^0_\alpha$, we obtain
\begin{eqnarray}
T^t_t= \rho,
\end{eqnarray}
\begin{eqnarray}
  T^i_i=-P_i=-p + 2 \eta \, \sigma^i_i,\,\, (i=x,y,z),
\end{eqnarray}
where there is no sum over $i$, and $\xi=0$. After adding the
inequalities $P_i \leq \rho$ ($i=x,y,z$), one obtains $p \leq
\rho$. On the other hand, after adding $-\rho \leq P_x $
($i=x,y,z$) we get $-\rho \leq p$. In both cases we have used the
result $\sigma^1_1+\sigma^2_2+\sigma^3_3=0$. This result is easy
generalized to the case where the bulk viscosity is taken into
account. Here, we obtain $-\rho+\theta \xi \leq p \leq \rho+\theta
\xi$.

\section{Scalar-tensor theories and the field 
equations}
We start with the Lagrangian density for the scalar tensor theory
of gravity
\begin{eqnarray}
L=\sqrt{-g}\left[ \phi (R-2 \Lambda(\phi))+\frac{\omega(\phi)
}{\phi }g^{\mu \nu }\partial _{\mu }\phi \partial _{\nu }\phi
\right],
\end{eqnarray}
where $g=det(g_{\alpha \beta})$. The arbitrary functions $\omega
=\omega \left( \phi \right)$ and $\Lambda =\Lambda \left( \phi
\right)$ distinguish the different scalar-tensor gravitational
theories: $\omega(\phi)$ is the coupling function and
$\Lambda(\phi)$ is a potential function and plays the role of a
cosmological constant .

The Euler-Lagrange equations of motion for $g_{\alpha \beta}$ and
$\phi$ obtained from the action $S= \int (L+L_{_M}) \, d^4x/16
\pi$ are
\begin{eqnarray}  \label{einstein-scalar}
R_{\alpha \beta}-\frac{1}{2} \, g_{\alpha \beta} R= - \frac{8
\pi}{\phi} T_{\alpha \beta} -\Lambda(\phi) g_{\alpha \beta}-
\nonumber \\ \frac{\omega(\phi)}{\phi^2} \left[ \phi_{,\alpha}
\phi_{,\beta}- \frac{1}{2}
g_{\alpha \beta} \phi_{,\gamma} \phi^{,\gamma} \right] - \frac{1}{\phi} %
\left[\phi_{;\alpha;\beta}-g_{\alpha \beta}
\phi^{;\gamma}_{\,\,\,\, ;\gamma} \right] ,
\end{eqnarray}
and
\begin{eqnarray}  \label{campo escalar}
\phi^{;\gamma}_{\,\,\,\, ;\gamma}+\frac{2 \phi^2 \frac{d
\Lambda(\phi)}{d \phi}-2 \phi \Lambda(\phi)}{2 \omega(\phi)+3} =
\frac{1}{2 \omega(\phi)+3} \times \nonumber \\ \left[ 8 \pi
T^{\gamma}_{\,\,\,\, \gamma}-\frac{d\omega(\phi)}{d \phi} \,
\phi_{,\gamma}\phi^{,\gamma} \right]
\end{eqnarray}
respectively, where the partial derivatives are denoted by a prime
and covariant derivatives are denoted by a semicolon. $T_{\alpha
\beta}$ is the stress energy-tensor which becomes calculated from
$L_{_M}$ through the definition $T_{\alpha
\beta}=\frac{2}{\sqrt{-g}} \frac{\delta}{\delta g_{\alpha \beta}}
\left [\sqrt{-g} \, L_{_M}  \right] $. General relativity is
recovered in the limit $\omega \longrightarrow \infty$ and
$\phi=const. \equiv 1/G$. We can conveniently
write~(\ref{einstein-scalar}) as
\begin{eqnarray}  \label{einstein-escalar simplificada}
R_{\alpha \beta}=\left[ \Lambda(\phi)-\frac{1}{2 \phi} \phi^{;
\gamma}_{\,\, ; \gamma} \right] g_{\alpha \beta}-\frac{8
\pi}{\phi} \left[T_{\alpha \beta} - \frac{1}{2} g_{\alpha \beta} T
\right]  \nonumber \\ -\frac{1}{\phi} \left[\phi_{;\alpha;\beta}+
\frac{\omega(\phi)}{\phi} \phi_{,\alpha} \phi_{,\beta} \right].
\end{eqnarray}

We now apply this theory to a homogeneous and anisotropic
cosmology of the Kasner type, expressed by the
metric~(\ref{Metric}), together with the energy-momentum tensor
specified by eq.~(\ref{tensor con shear}). The signature here is
($+ - - -$), and the Ricci tensor is given by $R_{\alpha
\beta}=R^{\gamma}_{\alpha \beta \gamma}$. The scalar expansion and
the traceless shear tensor are
\begin{eqnarray*}
\theta= S \, t^{-1},
\end{eqnarray*}
\begin{eqnarray*}
\sigma_{00}=0, \sigma_{ii}=t^{2p_i-1}\left [ \frac{S}{3}-p_i
\right],
\end{eqnarray*}
with
\begin{eqnarray*}
\sigma^2=\frac{1}{2 t^2} \left[ Q-\frac{S^2}{3} \right],
\end{eqnarray*}
respectively. Then, eq.~(\ref{campo escalar}) reduces to the
following field equations:
\begin{eqnarray}  \label{scalar field}
\ddot{\phi}+\frac{S}{t} \, \dot{\phi}+\frac{2 \phi^2
\Lambda^{'}(\phi)-2 \phi\, \Lambda(\phi)}{2 \omega(\phi)+3} =
\nonumber \\ \frac{8 \pi}{2 \omega(\phi)+3}
\left(\rho-3\left[p-\frac{S}{t} \, \xi \right]
  \right)- \frac{\omega^{\,'}}{2 \omega(\phi)+3%
}\, \dot{\phi}^2.
\end{eqnarray}
Eq.~(\ref{einstein-escalar simplificada}) gives rise to the
following set of eqs.:
\begin{eqnarray}  \label{cero-cero}
-\frac{S-Q}{t^2}=\Lambda(\phi)-\frac{8 \pi}{\phi}\left[\rho-\frac{1}{2}%
\left(\rho-3 \left[p-\frac{S}{t} \, \xi \right] \right) \right]
\nonumber
\\
-\frac{1}{2 \phi} \left[\ddot{\phi}+ \frac{S}{t} \dot{\phi} \right]-\frac{1}{%
\phi} \left[\ddot{\phi}+ \frac{\omega(\phi)}{\phi} \,
\dot{\phi}^2\right],
\end{eqnarray}
and
\begin{eqnarray}  \label{i-i}
p_i \left[1-S-\frac{16 \pi}{\phi} \, \eta \, t-\frac{\dot{\phi}}{\phi} \, t %
\right] t^{-2} =-\Lambda(\phi)-  \nonumber \\ \frac{4 \pi}{\phi}
\left[\rho- \left(p-\frac{S}{t} \, \xi \right)+\frac{4 S}{3 t} \,
\eta \right]+ \frac{1}{2 \phi} \left [ \ddot{\phi} + \frac{S}{t} \
\dot{\phi} \right],
\end{eqnarray}
where the dots specify derivatives with respect to the
cosmological time.

Since the $p_i$ ($i=1,2,3$) are all linearly independent we get
from eq.~(\ref{i-i}) that
\begin{eqnarray}  \label{eta}
\eta =\frac{\phi}{16 \pi \, t} \left( 1-S-\frac{\dot{\phi}}{\phi}
\, t \right) > 0.
\end{eqnarray}

 From the same eq., we get a condition that, after substituting eq.~(%
\ref{eta}) into this expression we get
\begin{eqnarray}  \label{condicion eta}
\frac{8 \pi \left(\rho-\left[p-\frac{S}{t} \, \xi \right]
\right)}{2 \phi} =\frac{1}{2 \phi} \left[\ddot{\phi} + \frac{5 S
\dot{\phi}}{3 t} \right] \nonumber \\ - \frac{S(1-S)}{3
t^2}-\Lambda(\phi).
\end{eqnarray}

This latter equation, together with eq.~(\ref{cero-cero}), allows
us to obtain explicit expressions for $\rho$ and $p$ which result
to be given by
\begin{eqnarray}  \label{scalar densidad}
8 \pi \, \rho= \frac{(S^2-Q) \, \phi}{2 t^2}+ \frac{S \dot{\phi}}{t} - \frac{%
\omega(\phi) \dot{\phi}^2}{2 \phi} - \phi \Lambda (\phi)
\end{eqnarray}
and
\begin{eqnarray}  \label{scalar presion}
8 \pi p= \frac{(S-Q) \, \phi}{2 t^2} - \ddot{\phi}- \frac{2 \dot{\phi} S}{3 t%
} + \frac{S(1-S)}{6 t^2} \, \phi -  \nonumber \\
\frac{\omega(\phi) \, \ddot{\phi}}{2 \phi} + \phi \Lambda(\phi)+8
\pi \frac{S}{t} \, \xi .
\end{eqnarray}

These two expressions can be substituted into the equation of
motion for the scalar field $\phi$, eq.~(\ref{scalar field}), and
obtain
\begin{eqnarray}  \label{scalar field 2}
\ddot{\phi}+\frac{S}{t}\,\dot{\phi}+\frac{2\phi ^{2}\Lambda
^{'}(\phi )-2\phi \,\Lambda (\phi )}{2\omega (\phi )+3} =
\frac{1}{(2\omega (\phi )+3)} \nonumber \\ \times \left[
\frac{\left( S^{2}+Q-2S\right) \phi }{t^{2}}+3\stackrel{\cdot
\cdot }{\phi }+\frac{3S}{t}\stackrel{\cdot }{\phi }+\frac{\omega \stackrel{%
\cdot }{\phi }^{2}}{\phi }-4\Lambda \phi \right] \\ - \frac{\omega
^{\, '}}{2\omega (\phi )+3}\,\dot{\phi}^{2}.  \nonumber
\end{eqnarray}
We should note that, when $\phi = const.= 1/G$ and $\Lambda(\phi)
= 0 $, this set of equations coincides with that considered by the
authors of ref\cite{Brevik}. In the following we shall consider
some particular cases.

\section{Some specific 
models}

The set of equations described above can be solved  for the
particular situation $\omega \left(
\phi \right) =\omega _{0}$ and $\Lambda (\phi )=\Lambda _{0}\phi ^{-\frac{2}{%
n}}$, where $\omega_0$, $\Lambda_0$ and $n$ are arbitrary real
constants. This choice allows us to write the following particular
solution for the scalar field: $\displaystyle \phi =\phi
_{0}\left(\frac{t}{t_0} \right) ^{n}$ as can see from~(\ref{scalar
field 2}). Here $\phi_0$ and $t_0$ are two constants that would be
associated with the present values of $\phi$ and $t$,
respectively. Replacing these expressions into eq.~(\ref{scalar
field 2}) we obtain the following expression for the $\Lambda_0$
parameter:
\begin{eqnarray}  \label{condicion constantes}
\displaystyle \Lambda_0 = \frac{n\,\phi_0^{2/n}}{2n-4} \left[
\frac{}{} n \, \omega_0(2-2S-n)-2S+S^2+Q \right].
\end{eqnarray}
These conditions give for $\rho$, $p$ and $\eta$ the following
expressions:
\begin{eqnarray}
\rho= \frac{\phi_0 \, t^{n-2}}{8 \pi(2-n)} \left[\frac{}{} S^2-Q+n
Q-n^2 S+n S-n^2 \omega_0S \right],
\end{eqnarray}
\begin{eqnarray}
p= \frac{\phi_0 \, t^{n-2}}{24 \pi(2-n)} \left[-3Sn -n
S^2+4S-S^2-3Q+3n^3 \right.  \nonumber \\
\left. -9n^2+2Sn^2+6n+3\omega_0 n^3+3\omega_0 n^2 S-6\omega_0n^2 \frac{}{} %
\right] \nonumber \\ +\frac{S}{t} \, \xi,
\end{eqnarray}
and
\begin{eqnarray}
\label{B-D eta} \eta=\frac{\phi_0}{16 \pi} \left[ \frac{}{} 1-S-n
\right] t^{n-1}.
\end{eqnarray}
We should note that  the set of values  $(\omega_0,n,\Lambda_0)$
will define interesting cases, which we consider in the following.
\begin{center}
{\bf A. Einstein Kasner type universe}
\end{center}
Einstein theory is obtained for $n=0$, $\omega_0 \longrightarrow
\infty$ and $\phi_0=\frac{1}{G}$. In this case the solutions take
the form
\begin{eqnarray}
\label{rho Einstein} \rho= \frac{t^{-2}}{16 \pi G} \left[\frac{}{}
S^2-Q \right],
\end{eqnarray}
\begin{eqnarray}
\label{p Einstein} p= \frac{t^{-2}}{48 \pi G} \left[4S-S^2-3Q
\frac{}{} \right] +\frac{S}{t} \, \xi,
\end{eqnarray}
and
\begin{eqnarray}
\label{eta Einstein} \eta=\frac{t^{-1}}{16 \pi G} \left[ \frac{}{}
1-S\right].
\end{eqnarray}

The corresponding particular cases are studied in detail in
ref.~\cite{Brevik}. We have shown above that the shear coefficient
of viscosity is always negative. As we saw earlier, the DEC imply
that $-\rho+\theta \xi \leq p \leq \rho+\theta \xi$. Then,
from~(\ref{rho Einstein}), (\ref{p Einstein}) and~(\ref{eta
Einstein}) for $p \leq \rho+\theta \xi$ we have that $1-S \leq 0$
and from $-\rho+\theta \xi \leq p$ we obtain that $-S^2-2 S+3Q
\leq 0$. Thus, from these expressions it is easy to see that one
always gets that $\eta \leq 0$ and $\rho \geq 0$.

\begin{center}
{\bf B. Brans-Dicke in a Kasner-type universe}
\end{center}

In this subsection we consider the Brans-Dicke theory, i.e. we
take $\Lambda_0=0$ with $n$ and $\omega_0$ for the moment
arbitrary parameters. Thus, from~(\ref{condicion constantes}), we
have that the shear viscosity is still given by~(\ref{B-D eta})
and the energy density and pressure are given by
\begin{eqnarray}  \label{BDrho}
\kappa \rho= \frac{\phi_0 t^{n-2}}{2} \left[\frac{}{} S^2+2Sn-Q-\omega_0 n^2 %
\right]
\end{eqnarray}
and
\begin{eqnarray}  \label{BDp}
\kappa p= \frac{\phi_0 t^{n-2}}{6} \left[\frac{}{}
4S-S^2-3Q-n(6n-6+4S) \right.  \nonumber \\ \left. -3\omega_0 n^2
\frac{}{} \right],
\end{eqnarray}
respectively.

 From eqs.~(\ref{BDrho}) and~(\ref{BDp}) we see that a barotropic
state equation is satisfied ($p= \gamma \rho$), and then
\begin{eqnarray}  \label{BDgamma}
\gamma = \frac{4S-S^2-3Q-n(6n-6+4S)-3\omega_0
n^2}{3(S^2+2Sn-Q-\omega_0 n^2)}
\end{eqnarray}
and the constraint for the constants is
\begin{eqnarray}  \label{constraint for the constants}
2 n \omega_0(1-S-n/2)=2S-S^2-Q,
\end{eqnarray}
which is obtained from eq.~(\ref{condicion constantes}). This
solution must satisfy the second law of thermodynamics ($\eta \geq
0$) and the DEC ($\rho \geq 0$ and $-1 \leq\gamma \leq 1$%
), i.e. there are four conditions to be satisfied. It is easy to
show that the three inequalities of the DEC are linearly
dependent. Then we get from $\rho \geq 0$
\begin{eqnarray}  \label{BDrhocondition}
S^2+2Sn-Q-\omega_0 n^2 \geq 0
\end{eqnarray}
and from $p \leq \rho$
\begin{eqnarray}  \label{BDprhocondition}
4S^2-4S+6n^2-6n+10nS \geq 0.
\end{eqnarray}

Now, from the conditions $\eta \geq 0$ and~(\ref{BDrhocondition})
we obtain
\begin{eqnarray}  \label{oneBDcondition}
S(1+n)\geq Q+\omega_0 n^2.
\end{eqnarray}
Thus, a physically consistent model must satisfy the
conditions~(\ref{constraint for the constants}), (\ref
{BDprhocondition}), and~(\ref{oneBDcondition}).

It can be shown that the condition~(\ref{constraint for the
constants}) is compatible with the
inequality~(\ref{oneBDcondition}), but not with the
condition~(\ref {BDprhocondition}). Therefore,  in this particular
model based on  Brans-Dicke theory, there are no physically
reasonable solutions for a viscous Kasner type model. This is
similar to what is found in  Einstein's theory.

\begin{center}
{\bf C. Scalar-tensor Kasner type universe}
\end{center}

In the following we do not consider the bulk viscosity, since as
is well known, the shear viscosity at early time is much greater
than the bulk viscosity~\cite{Caderni}; therefore, we could  take
$\xi=0$. In this case we can write a barotropic equation of state
$p= \gamma \rho$, where, in agreement with the DEC, $\gamma$ lies
in the interval $-1\leq \gamma \leq 1$.

It is easy to check that in this case there is a family of
solutions which satisfy the second law of thermodynamics together
with  the DEC, simultaneously. For instance, we have found the
following particular solutions. When the parameters $\omega$ and
$n$ take the values $\omega_0=0.1$ and  $n=-0.1$, it is found that
the parameters  $p_1=0.1$, $p_2=0.12$ and $p_3=0.8$ represent a
solution of the field equations. Here, all the relevant
quantities, such that the effective pressure $p$, the shear
viscosity $\eta$, the energy density $\rho$, etc. all these
quantities are physically acceptable. Thus, there are particular
situations in scalar-tensor theories of gravity where are not
found the problems appearing in Einstein theory.

Note that if $\eta=0$, i.e. $n=1-S$, one has the solution
\begin{eqnarray}
\hspace{-1.0cm} \rho=S p=\frac{\phi_0 t^{-(1+S)}}{8 \pi(1+S)} S
\hspace{2cm} \nonumber
\\ \times \left(2 \omega_0 S-\omega_0-\omega_0 S^2+2S-S^2-Q
\right),
\end{eqnarray}
from which we get the barotropic equation of state, $p=\frac{1}{S}
\, \rho$, with $-1 \leq 1/S\leq 1$. A dust dominated universe is
obtained when $S \longrightarrow \infty$. If $S=1$ ($n=0$), then
the model becomes stiff matter dominated. In the latter case we
obtain
\begin{eqnarray}
\rho=p=\frac{\phi_0}{16 \pi t^2} \, (1-Q),
\end{eqnarray}
which coincides with equation (21) of ref.~\cite{Brevik}.

\section{Generation  of entropy}
In the following we shall determine the generation of entropy in a
scalar-tensor theory. Here we include the shear coefficient of
viscosity only, since $\xi \ll \eta$.

It is well known that the production of entropy could be related
to the anisotropy of the universe\cite{Mi}. In order to see this
we introduce the entropy current four-vector $S^{\mu}$ as
\begin{equation}
S^{\mu} = n_{_{b}} k_{_{B}} \sigma u^{\mu},
\end{equation}
where, as before, $u^{\mu}$ represents the four-velocity,
$n_{_{b}}$ the baryon number density, $k_{_{B}}$ is the Boltzmann
constant and $\sigma$ the nondimensional entropy per baryon. Since
\begin{eqnarray}
\label{generacion de entropia} S^{\mu}_{\,\, ; \mu}=\frac{2}{T} \,
\eta \, \sigma^{\mu \nu} \sigma_{\mu \nu}.
\end{eqnarray}
($\xi=0$), it is obtained that\cite{Brevik1}
\begin{eqnarray}
\label{sigmapunto} \dot{\sigma}=\frac{2 S^2}{n_{_B} k_B T t^2} \,
\eta\, A,
\end{eqnarray}
where $\displaystyle A=\frac{3 Q}{S^2}-1$.

We note that from~(\ref{B-D eta}) one has $\eta \, t^{1-n}=
const$. This fact permits us to write all the quantities if we
know the value of $\eta$ at a given moment of time. We use here
the values given in ref~\cite{Brevik} for some parameters at
$t=1000$ s (the universe is characterized by ionized H and He in
approximate equilibrium with radiation). At that time, the number
densities of electrons and protons,
$n_{e_{_{1000}}}=n_{p_{_{1000}}} \simeq 10^{19} \, cm^{-3}$, the
temperature $T_{_{1000}} \simeq 4 \times 10^{8}$ K and
$\eta_{_{1000}} \simeq 2.8 \times 10^{14} \ g \, cm^{-1} s^{-1}$.
Then we get at any time $t$
\begin{eqnarray}
\eta(t)= \eta_{_{1000}} \, \left( \frac{1000}{t} \right)^{1-n}.
\end{eqnarray}
Note that from~(\ref{sigmapunto}) we have $\displaystyle
\frac{n_{_B} \, T \, t^2 \, \dot{\sigma}}{\eta}=const$. Thus the
generation of entropy at any time can be written as
\begin{eqnarray}
\dot{\sigma}(t) = \dot{\sigma}_{_{1000}} \,
\frac{n_{e_{_{1000}}}}{n_e(t)} \frac{T_{_{1000}}}{T(t)}
\left(\frac{1000}{t}\right)^{3-n}.
\end{eqnarray}

In order to determine how the generation of entropy decreases as
time goes on, we determine this generation of entropy for two
different instants in  the evolution of the universe. We consider
the plasma era (radiation era), where the temperature ranges
$10^{10} \, K \geq T \geq 4000 \, K$~\cite{Caderni}. For the
recombination of hydrogen we have $t_{_{rec}} \simeq 0.126 \times
10^{14}$, $n_{e_{_{rec}}}=n_{p_{_{rec}}} \simeq 4 \times 10^{3} \,
cm^{-3}$, the temperature $T_{_{rec}} \simeq 4000$ K. Then, for
the ratio between values of generation of entropy at time of
recombination $t_{_{rec}}$ and the time $t=1000$ we get
for the model studied above ($n=-0.1$), that $\displaystyle
\frac{\dot{\sigma}_{_{rec}}}{\dot{\sigma}_{_{1000}}} \simeq
10^{-11}$, which implies a drastic reduction in the rate of
entropy production during the plasma era.

\section{Conclusions}
We have studied the scalar-tensor theories for Kasner type metrics
in the presence of a viscous fluid. We have explored the
possibility of describing reasonable physical models for a Kasner
type universe when we keep the thermodynamics ($\eta \geq 0$ and
$\xi \geq 0$) together with the dominant energy condition in
theories more general than Einstein's gravity. We concluded, for
the specific model treated here, that in the Brans-Dicke theory,
like in the Einstein theory, it is not possible to describe the
growth of entropy.

In the case of the scalar-tensor theory we have found, for our
particular choice of the scalar field $\phi$, the coupling
function $\omega(\phi)=\omega_0$ and the potential function
$\Lambda(\phi)=\Lambda_0 \phi^{-2/n}$, that it is possible to keep
the thermodynamic conditions ($\eta \geq 0$ and $\xi \geq 0$)
together with the DEC, as was described in Sect.IV.C. For these
models we have found that, during the plasma era, i.e. from $T\sim
10^{10}$ K to $T_{rec}\sim 4.000$ K, a drastic reduction of the
rate of entropy occurred.

\section{Acknowledgements}

We thank Paul Minning for carefully reading the manuscript. This
work was supported by COMISION NACIONAL DE CIENCIAS Y TECNOLOGIA
through Grants FONDECYT N$^{0}$ 1990601 (MC and PS) and N$^{0}$
1000305 (SdC). Also MC was supported by Direcci\'{o}n de
Promoci\'{o}n y Desarrollo de la Universidad del B\'\i o-B\'\i o,
SdC was supported from UCV-DGIP 123.752/2000 and PS was supported
in part by Direcci\'{o}n de Investigaci\'{o}n de la Universidad de
Concepci\'{o}n through Grant N 98.011.023-1.0.


\begin{references}
\bibitem{Guth}  A.H. Guth, Phys. Rev. D 23 (1981) 347; A.D. Linde, Phys. 
Lett. B 108 (1982) 389; A.
Albrecht, P.J. Steinhardt, Phys. Rev. D 48 (1982) 1220;  A.D.
Linde, Phys. Lett. B 129 (1983) 177.


\bibitem{Weinberg}  S. Weinberg, Gravitation and Cosmology (Wiley, New York,
1979) pp 593-594.


\bibitem{Johri}  V.B. Johri and R. Sudharsan Phys. Lett. A 132 (1988) 316.


\bibitem{Tupper1}  A.A. Coley and B.O.J. Tupper, Astrophy. J. (1984) 180 26.


\bibitem{Barrow}  J.D. Barrow, Phys. Rev. D 55 (1997) 7451.


\bibitem{Cataldo1}  M. Cataldo and Sergio del Campo, Phys. Rev. D 61 (2000)
128301.


\bibitem{Brevik1}  I. Brevik and S.V. Pettersen, Phys. Rev. D 61
(2000)127305.


\bibitem{Brevik}  I. Brevik, S.V. Pettersen, Phys. Rev. D 56 (1997) 3322.


\bibitem{MiThWe}  C. W. Misner, K. S. Thorner and J. A. Wheeler, {\it %
Gravitation} (Freeman and Co., San Francisco, 1973).

\bibitem{Hawking}  S.W. Hawking and G.F.R. Ellis, The large scale structure
of space-time (Cambridge Monographs on Mathematical Physics, 1980)
\bibitem{Caderni} N. Caderni, R. Fabbri 1977 Phys. Lett. B {\bf 69}
508.
\bibitem{Mi} C. W. Misner 1968 Astrophy. J. {\bf 151} 431; S.
Weinberg 1971 Astrophy. J. {\bf 168} 175.
\end{references}
\end{document}